\documentstyle[stwol]{article}
\input{psfig}

\def\beq{\begin{equation}}
\def\eeq{\end{equation}}
\def\eq{\end{equation}}
\def\to{\rightarrow}

\def\bsg{\ifmmode B\to X_s\gamma\else $B\to X_s\gamma$\fi}
\def\bsll{\ifmmode B\to X_s\ell^+\ell^-\else $B\to X_s\ell^+\ell^-$\fi}
\def\bstt{\ifmmode B\to X_s\tau^+\tau^-\else $B\to X_s\tau^+\tau^-$\fi}
\def\shat{\ifmmode \hat{s}\else $\hat{s}$\fi}

\newcommand{\newc}{\newcommand}

\newc{\lcal}{\int {\cal L}dt}

\newc{\mHpm}{m_{H^\pm}}
\newc{\gsim}{\lower.7ex\hbox{$\;\stackrel{\textstyle>}{\sim}\;$}}
\newc{\lsim}{\lower.7ex\hbox{$\;\stackrel{\textstyle<}{\sim}\;$}}
\newc{\ie}{{\it i.e.}}          
\newc{\etal}{{\it et al.}}
\newc{\eg}{{\it e.g.}}          
\newc{\kev}{\hbox{\rm\,keV}}            
\newc{\mev}{\hbox{\rm\,MeV}}            
\newc{\gev}{\hbox{\rm\,GeV}}            
\newc{\tev}{\hbox{\rm\,TeV}}
\newc{\xpb}{\hbox{\rm\, pb}}
\newc{\xfb}{\hbox{\rm\, fb}}

\newc{\mtop}{m_t}
\newc{\mbot}{m_b}
\newc{\mz}{m_Z}
\newc{\mw}{M_W}
\newc{\alphasmz}{\alpha_s(m_Z^2)}
\newc{\swsq}{\sin^2\theta_W}
\newc{\tw}{\tan\theta_W}
\newc{\cw}{\cos\theta_W}
\newc{\sw}{\sin\theta_W}
\newc{\BR}{\hbox{\rm BR}}
\newc{\zbb}{Z\to b\bar}
\newc{\Gb}{\Gamma (Z\to b\bar b)}
\newc{\Gh}{\Gamma (Z\to \hbox{\rm hadrons})}
\newc{\rbsm}{R_b^\hbox{\rm sm}}
\newc{\rbsusy}{R_b^\hbox{\rm susy}}
\newc{\drb}{\delta R_b}

\newc{\sgn}{\mbox{sgn}}
\newc{\tbeta}{\tan\beta}
\newc{\uL}{{\tilde u_L}}
\newc{\uR}{{\tilde u_R}}
\newc{\cL}{{\tilde c_L}}
\newc{\cR}{{\tilde c_R}}
\newc{\tL}{{\tilde t_L}}
\newc{\tR}{{\tilde t_R}}
\newc{\dL}{{\tilde d_L}}
\newc{\dR}{{\tilde d_R}}
\newc{\sL}{{\tilde s_L}}
\newc{\sR}{{\tilde s_R}}
\newc{\bL}{{\tilde b_L}}
\newc{\bR}{{\tilde b_R}}
\newc{\eL}{{\tilde e_L}}
\newc{\eR}{{\tilde e_R}}
\newc{\mhp}{m_{H^\pm}}
\newc{\mhalf}{m_{1/2}}

\newc{\lR}{\tilde{l}_R}
\newc{\lL}{\tilde{l}_L}
\newc{\nL}{\tilde{\nu}_L}
\newc{\na}{\chi^0_1}
\newc{\nb}{\chi^0_2}
\newc{\nc}{\chi^0_3}
\newc{\nd}{\chi^0_4}
\newc{\ca}{\chi^{\pm}_1}
\newc{\cb}{\chi^{\pm}_2}
\newc{\camp}{\chi^\mp_1}
\newc{\cbmp}{\chi^\mp_1}
\newc{\capos}{\chi^{+}_1}
\newc{\caneg}{\chi^{-}_1}

\def\MPL #1 #2 #3 {{\em Mod. Phys. Lett.} {\bf#1},\ #2 (#3)}
\def\NPB #1 #2 #3 {{\em Nucl. Phys.} {\bf#1},\ #2 (#3)}
\def\PLB #1 #2 #3 {{\em Phys. Lett.} {\bf#1},\ #2 (#3)}
\def\PR #1 #2 #3 {{\em Phys. Rep.} {\bf#1},\ #2 (#3)}
\def\PRD #1 #2 #3 {{\em Phys. Rev.} {\bf#1},\ #2 (#3)}
\def\PRL #1 #2 #3 {{\em Phys. Rev. Lett.} {\bf#1},\ #2 (#3)}
\def\RMP #1 #2 #3 {{\em Rev. Mod. Phys.} {\bf#1},\ #2 (#3)}
\def\ZPC #1 #2 #3 {{\em Z. Phys.} {\bf#1},\ #2 (#3)}
\def\beq{\begin{equation}}
\def\eeq{\end{equation}}
\def\bea{\begin{eqnarray*}}
\def\eea{\end{eqnarray*}}
\def\slashchar#1{\setbox0=\hbox{$#1$}           
   \dimen0=\wd0                                 
   \setbox1=\hbox{/} \dimen1=\wd1               
   \ifdim\dimen0>\dimen1                        
      \rlap{\hbox to \dimen0{\hfil/\hfil}}      
      #1                                        
   \else                                        
      \rlap{\hbox to \dimen1{\hfil$#1$\hfil}}   
      /                                         
   \fi}                                         %
\catcode`@=11
\long\def\@caption#1[#2]#3{\par\addcontentsline{\csname
  ext@#1\endcsname}{#1}{\protect\numberline{\csname
  the#1\endcsname}{\ignorespaces #2}}\begingroup
    \small
    \@parboxrestore
    \@makecaption{\csname fnum@#1\endcsname}{\ignorespaces #3}\par
  \endgroup}
\catcode`@=12

\begin{document}
\renewcommand{\thefootnote}{\fnsymbol{footnote}} 
\def\srf#1{$^{#1}$\ }
\def\mainhead#1{\setcounter{equation}{0}\addtocounter{section}{1}
  \vbox{\begin{center}\large\bf #1\end{center}}\nobreak\par}
\def\subhead#1{\vbox{\smallskip\noindent \bf #1}\nobreak\par}
\def\autolabel#1{\auto\label{#1}}
\hfuzz=1pt
\renewcommand{\arraystretch}{1.5}

\begin{titlepage} 
\rightline{\vbox{\halign{&#\hfil\cr
&SLAC-PUB-7338\cr
&October 1996\cr}}}
\vspace{1in} 
\begin{center}

{\Large\bf
PROBING SUPERSYMMETRY IN RARE B DECAYS}
\footnote{Work supported by the Department of 
Energy, Contract DE-AC03-76SF00515}
\medskip

\normalsize 
{\large J.L. Hewett}\footnote{Work performed in collaboration with J.D. Wells}
\vskip .3cm
Stanford Linear Accelerator Center \\
Stanford CA 94309, USA\\
\vskip .3cm

\end{center}

\begin{abstract} 

We determine the ability of future experiments to observe supersymmetric
contributions to the rare decays $B \to X_s \gamma$ and 
$B \to X_s l^+l^-$.  A global fit to the Wilson coefficients which contribute 
to these decays is performed from Monte Carlo generated data.  This fit is
then compared to supersymmetric predictions for
several different patterns of the superpartner spectrum.

\end{abstract} 

\vskip1.75in
\noindent{(Talk given at the {\it 28th International Conference on High
Energy Physics}, Warsaw, Poland, July 1996.)}

\renewcommand{\thefootnote}{\arabic{footnote}} \end{titlepage} 

\newpage

\title{PROBING SUPERSYMMETRY IN RARE B DECAYS}

\author{ J.L. Hewett}

\address{Stanford Linear Accelerator Center,
Stanford University, Stanford, CA 94309
}


\twocolumn[\maketitle\abstracts{
We determine the ability of future experiments to observe supersymmetric
contributions to the rare decays $B \to X_s \gamma$ and 
$B \to X_s l^+l^-$.  A global fit to the Wilson coefficients which contribute 
to these decays is performed from Monte Carlo generated data.  This fit is
then compared to supersymmetric predictions for
several different patterns of the superpartner spectrum.
}]


The first conclusive observation of penguin mediated processes, the
exclusive $B\to K^*\gamma$ and inclusive \bsg, by CLEO \cite{cleo:94}
has placed the study of rare $B$ decays on new ground.  These flavor
changing neutral current (FCNC) transitions provide an essential opportunity
to test the Standard Model (SM) and offer a complementary strategy in
the search for new physics by probing the indirect effects of new particles
and interactions in higher-order processes.  
With the expected high luminosity of the $B$-Factories 
presently under construction (and the associated advanced detector technology),
radiative $B$ decays will no longer be rare 
events, and the exploration of FCNC transitions can continue by probing
decay modes with even smaller predicted branching fractions.  The cleanest
rare decay which occurs at a rate accessible to these 
machines is \bsll.  In fact, experiments at $e^+e^-$
and hadron colliders are already closing in on the observation \cite{bsll} of 
the exclusive modes $B\to K^{(*)}\ell^+\ell^-$ with $\ell=e$ and  $\mu$, 
respectively.  Once this decay is observed, the utilization of the
kinematic distributions of the $\ell^+\ell^-$ pair, such as
the lepton pair invariant mass distribution and forward
backward asymmetry \cite{ali}, and the tau polarization 
asymmetry~\cite{hewett} 
in $B\to X_s\tau^+\tau^-$, together with $B(\bsg)$ will provide a stringent
test of the SM.  In this talk we determine the ability of $B$-Factories to
probe possible supersymmetric contributions to these decays.

Softly broken supersymmetry (SUSY) is a 
decoupling theory, thus making it a challenge to search for its effects
through indirect methods.  However, a promising
approach is to measure
observables where supersymmetry and the SM arise at the
same order in perturbation theory. In this case the SUSY
contributions do not suffer an extra $\alpha /4\pi$ reduction compared
to the SM amplitudes.  The relative ratio between the 
lowest order SM amplitudes and those of supersymmetry 
could then be ${\cal O}(1)$ if $\tilde m\simeq \mw$.
Rare $B$-decays could then provide such an opportunity for discovering
indirect effects of supersymmetry \cite{jimjo,rareb}.

The effective field theory for $b\to s$ transitions which incorporates 
QCD corrections is governed by the Hamiltonian
\begin{equation}
{\cal H}_{eff}={-4G_F\over\sqrt 2}V_{tb}V^*_{ts}\sum_{i=1}^{10}C_i(\mu)
{\cal O_i}(\mu)\,,
\label{effham}
\end{equation}
where the ${\cal O}_i$ are a complete set of renormalized operators of
dimension six or less which mediate $b\to s$ transitions and are 
catalogued in, \eg , Ref.~7.  The $C_i$ represent
the corresponding Wilson coefficients which are evaluated perturbatively
at the electroweak scale where the matching conditions are imposed and
then evolved down to the renormalization scale $\mu\approx  m_b$.  The
expressions for $C_i(M_W)$ are given by the Inami-Lim functions \cite{inami}.

For \bsll\ this formalism leads to the physical decay amplitude (neglecting 
$m_s$)
\begin{eqnarray}
{\cal M} &=& {\sqrt 2G_F\alpha\over \pi}V_{tb}V^*_{ts}
\left[ C_9^{eff}\bar s_L\gamma_\mu b_L\bar\ell\gamma^\mu\ell \right. \nonumber\\
& & \quad +C_{10}\bar s_L\gamma_\mu b_L\bar\ell\gamma^\mu\gamma_5\ell 
\nonumber\\
& & \quad\left.
-2C^{eff}_7m_b\bar s_Li\sigma_{\mu\nu}{q^\nu\over q^2}b_R\bar\ell\gamma^\mu\ell
\right] \,, 
\end{eqnarray}
where $q^2$ represents the momentum transferred to the lepton pair.  
We incorporate the NLO analysis
for this decay which has been performed in Buras \etal~\cite{buras:95}, where
it is stressed that a scheme independent result can only be obtained by 
including the leading and next-to-leading logarithmic corrections to
$C_9(\mu)$ while retaining
only the leading logarithms in the remaining Wilson coefficients.  The
residual leading $\mu$ dependence in $C_9(\mu)$ is cancelled by that
contained in the matrix element of ${\cal O}_9$, yielding an effective value 
$C^{eff}_9$.
The effective value for $C^{eff}_7(\mu )$ refers to the leading
order scheme independent result. 
The operator ${\cal O}_{10}$ does not renormalize.
The numerical estimates (in the naive dimensional regularization (NDR)
scheme) for these
coefficients are displayed in Table \ref{coeffval}.
The reduced scale dependence of the NLO versus the LO corrected coefficients
is reflected in the deviations $\Delta C_9(\mu)\lsim\pm 10\%$ and 
$\Delta C^{eff}_7(\mu)\approx\pm 20\%$ as $\mu$ is varied.
We find that the values of the coefficients are much less
sensitive to the remaining input parameters, with $\Delta
C_9(\mu),\Delta C^{eff}_7(\mu)\lsim 3\%$, varying 
$\alpha_s(M_Z)=0.118\pm 0.003$ \cite{pdg,schmelling}, 
and $m_t^{phys}=175\pm 6\gev$ \cite{cdfd0}.
The resulting inclusive branching fractions (which are computed by scaling
the width for \bsll\ to that for $B$ semi-leptonic decay)  are found
to be $(6.25^{+1.04}_{-0.93})\times 10^{-6}$,
$(5.73^{+0.75}_{-0.78})\times 10^{-6}$, 
and $(3.24^{+0.44}_{-0.54})\times 10^{-7}$
for $\ell=e, \mu$, and $\tau$, respectively, taking into account the
above input parameter ranges, as well as 
$B_{sl}\equiv B(B\to X\ell\nu)=(10.23\pm 0.39)\%$~\cite{richman}, and
$m_c/m_b=0.29\pm 0.02$.

\begin{table}\begin{center}
\caption{Values of the Wilson coefficients for several choices of the
renormalization scale.  Here, we take $m_b=4.87\gev$, 
$m_t=175\gev$, and $\alpha_s (M_Z)=0.118$.}
\label{coeffval}
\vspace{0.4cm}
\begin{tabular}{|c|c|c|c|} \hline
& $\mu=m_b/2$ & $\mu=m_b$ & $\mu=2m_b$ \\ \hline
$C^{eff}_7$  & $-0.371$ & $-0.312$ & $-0.278$ \\
$C_9$ & 4.52 & 4.21 & 3.81 \\
$C_{10}$ & $-4.55$ & $-4.55$ & $-4.55$ \\ \hline
\end{tabular}
\end{center}
\end{table}

The operator basis for the decay \bsg\ contains the first eight operators
in the effective Hamiltonian of Eq.~(\ref{effham}).  The leading logarithmic 
QCD corrections have been completely resummed, 
but lead to a sizeable $\mu$ dependence of the branching fraction 
and hence it is essential to include the next-to-leading order corrections.
In this case, the calculation 
involves several steps, requiring NLO
corrections to both $C_7^{eff}$ and the matrix element of ${\cal O}_7$.
For the matrix element, this includes the QCD bremsstrahlung 
corrections \cite{greub:91} $b\to s\gamma+g$, and the NLO virtual corrections
which have recently been completed \cite{greub:96}.  
Summing these contributions to the matrix elements
and expanding them around $\mu=m_b$, one arrives at the decay 
amplitude
\begin{equation}
{\cal M}(b\to s\gamma) = -{4G_FV_{tb}V^*_{ts}\over\sqrt 2}D\langle
s\gamma|{\cal O}_7(m_b)|b\rangle_{tree} \,, 
\end{equation}
with
\begin{eqnarray}
\label{dc7eq}
D=C_7^{eff}(\mu)&+&{\alpha_s(m_b)\over 4\pi}\left( C_i^{(0)eff}(\mu)
\gamma_{i7}^{(0)}\log {m_b\over\mu}\right. \nonumber\\
& & \left.  +\, C_i^{(0)eff}r_i\right) \,.
\end{eqnarray}
Here, the quantities $\gamma_{i7}^{(0)}$ are the entries of the effective
leading order anomalous dimension matrix, and the $r_i$ are computed in Greub
\etal~\cite{greub:96}, for $i=2,7,8$.  The first term in Eq.~\ref{dc7eq}, 
$C_7^{eff}(\mu)$, must be computed
at NLO precision, while it is consistent to use the leading order values
of the other coefficients.  
For $C_7^{eff}$ the NLO result entails
the computation of the ${\cal O}(\alpha_s)$ terms in the matching conditions
\cite{yao},
and the renormalization group evolution of $C_7^{eff}(\mu)$ must be computed
using the ${\cal O}(\alpha_s^2)$ anomalous dimension matrix.  
Preliminary NLO results for these anomalous dimensions have 
recently been reported~\cite{misiak:96}, with
the conclusion being that in the NDR scheme the NLO
correction to $C^{eff}_7(\mu)$ 
is small.  Therefore, a good approximation for the
inclusive width is obtained by employing the leading order expression for 
$C^{eff}_7(\mu )$, 
with the understanding that this introduces a small inherent 
uncertainty in the calculation.  We then find the branching fraction
(again, scaling to semi-leptonic decay)
\begin{equation}
B(\bsg)=(3.25\pm 0.30 \pm 0.40)\times 10^{-4} \,,
\end{equation}
where the first error corresponds to the combined uncertainty associated with 
the value of $m_t$ and $\mu$, and the second error represents the uncertainty
from $\alpha_s(M_Z), B_{sl}$, and $m_c/m_b$.
This is well within the range observed by CLEO \cite{cleo:94} which is
$B=(2.32\pm 0.57\pm 0.35)\times 10^{-4}$ with the $95\%$ C.L.
bounds of $1\times 10^{-4}< B(\bsg)<4.2\times 10^{-4}$.  

\begin{figure}
\centerline{
\psfig{figure=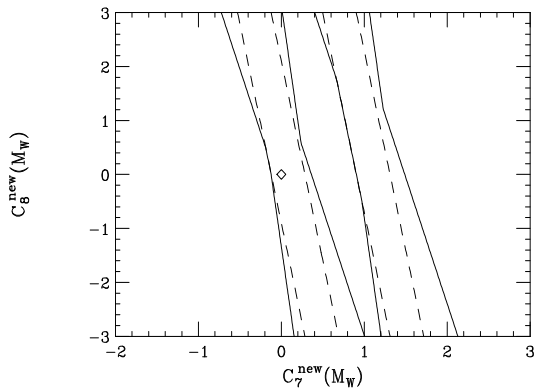,height=2.0in,width=2.75in}}
\vspace*{-0.25cm}
\caption{Bounds
on the contributions from new physics to $C_{7,8}$.  The region allowed
by the CLEO data corresponds to the area inside the solid diagonal bands.
The dashed bands represent the constraints when the renormalization scale
is set to $\mu =m_b$. 
The diamond at the position
(0,0) represents the standard model.}
\label{fig2}
\end{figure}

Measurements of $B(\bsg)$ alone constrain the magnitude, but not the sign,
of $C^{eff}_7(\mu)$.  We can write the coefficients
at the matching scale in the form $C_i(M_W)=C_i^{SM}(M_W)+C_i^{new}(M_W)$,
where $C_i^{new}(M_W)$ represents the contributions from new
interactions.  Due to operator mixing, \bsg\ then limits the 
possible values for $C_i^{new}(M_W)$ for $i=7,8$.  These bounds are
summarized in Fig.~\ref{fig2}.
Here, the solid bands correspond to the
constraints obtained from the current CLEO measurement,
taking into account the variation of the renormalization scale
$m_b/2 \leq \mu \leq 2m_b$, as well as the allowed ranges of
the other input parameters.  The dashed bands represent the constraints
when the scale is fixed to $\mu =m_b$.  We note that large values of
$C_8^{new}(\mw)$ are allowed even in the region where
$C_7^{new}(\mw)\simeq 0$.  

\begin{figure}[t]
\centerline{
\psfig{figure=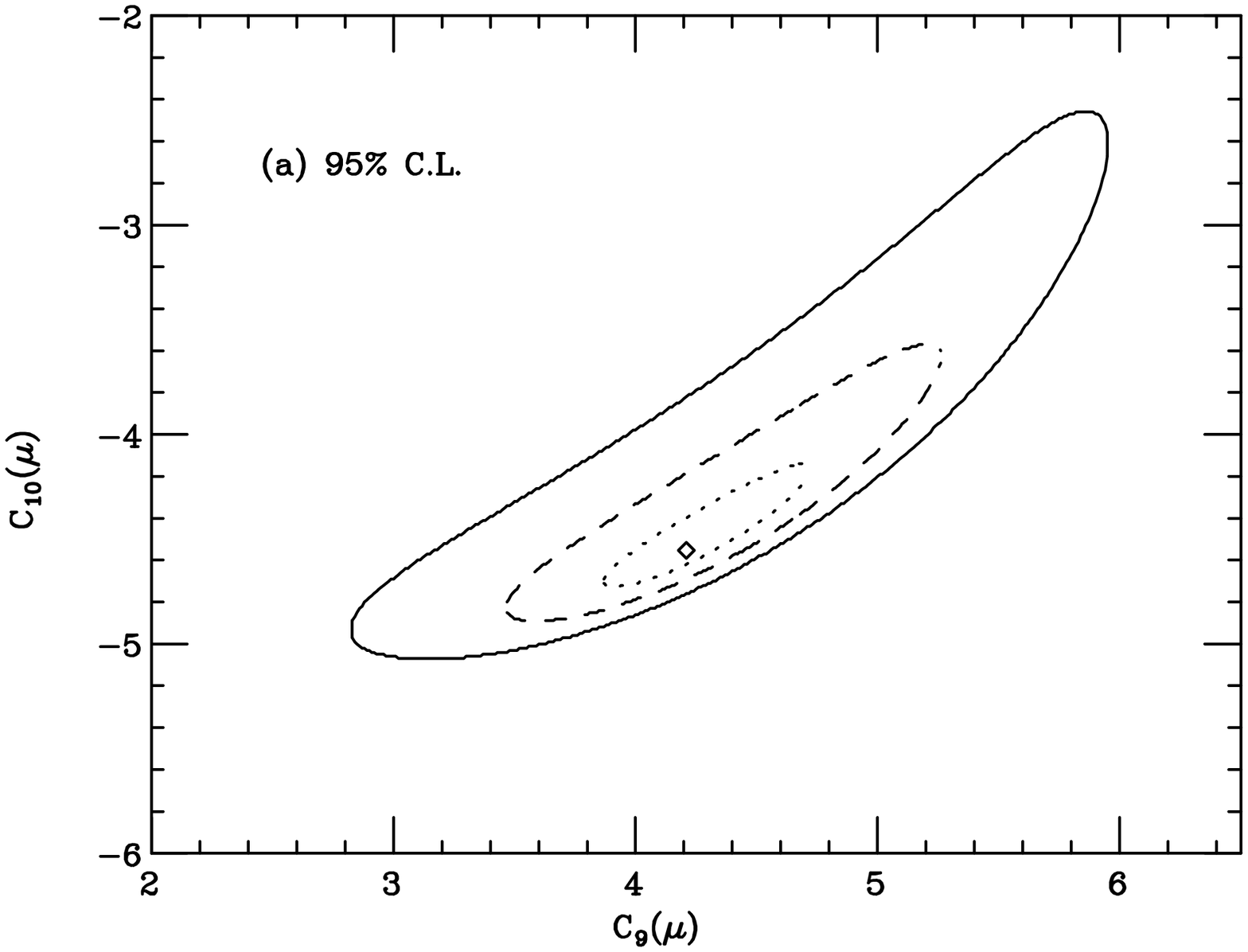,height=2.0in,width=2.75in,angle=-90}}
\vspace*{-0.75cm}
\centerline{
\psfig{figure=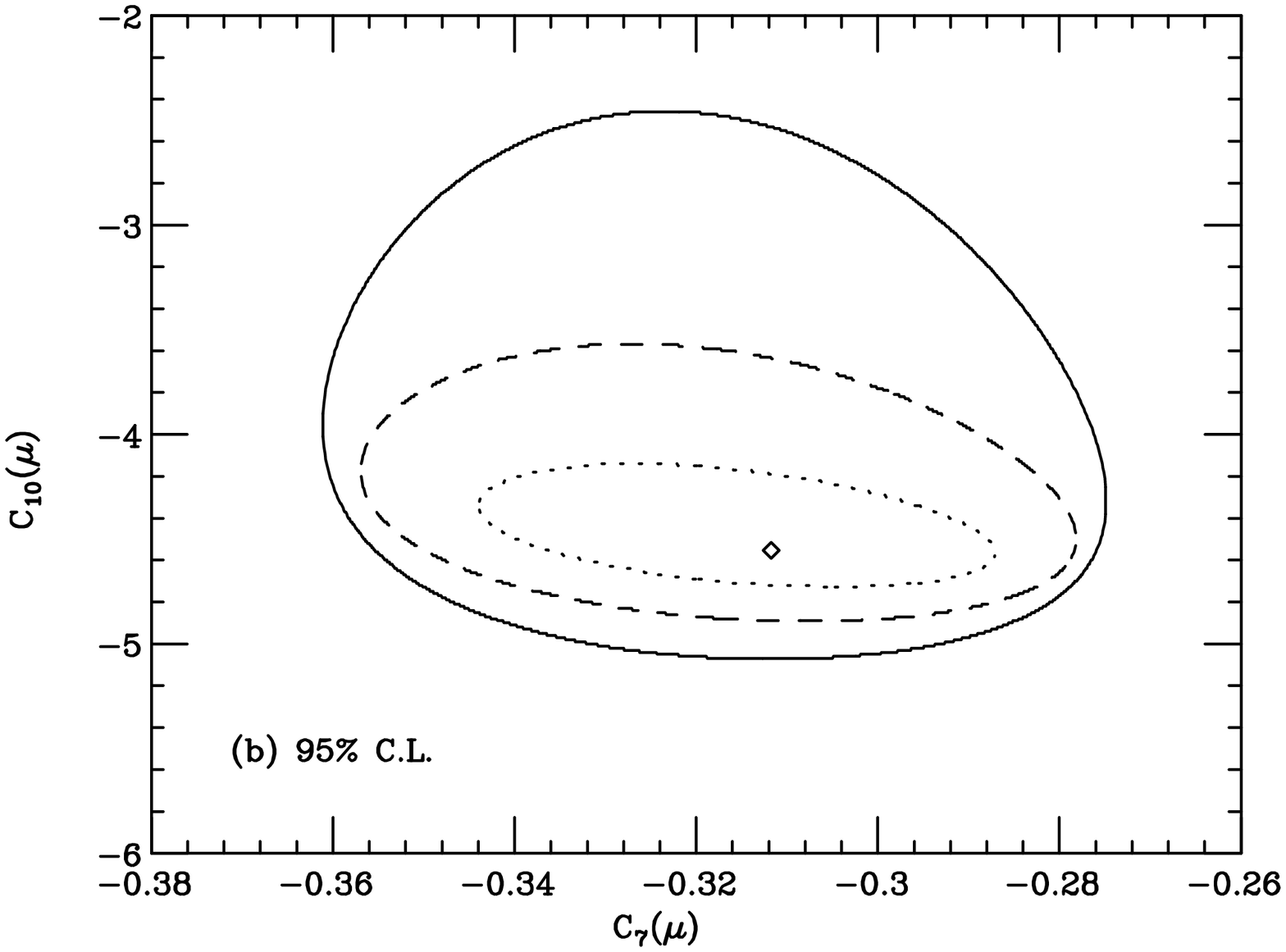,height=2.0in,width=2.75in,angle=-90}}
\vspace*{-0.75cm}
\caption{The $95\%$
C.L. projections in the (a) $C_9 - C_{10}$ and (b) $C^{eff}_7 - C_{10}$ 
planes, 
where the allowed regions lie inside of the contours.  The solid, dashed, and 
dotted contours correspond to $3\times 10^7$, $10^8$, and $5\times 10^8$
$B\bar B$ pairs. The central value of the 
SM prediction is labeled by the diamond.}
\label{fig3}
\end{figure}

Measurement of the kinematic distributions associated with the final
state lepton pair in \bsll\ as well as the rate for \bsg\ allows for
the determination of the sign and magnitude of all the Wilson coefficients
for the contributing operators in a model independent
fashion.  We have performed a Monte Carlo analysis in order to
ascertain how much quantitative information will be obtainable at
future $B$-factories and follow the procedure outlined in Ref. 5.
For the
process \bsll, we consider the lepton pair invariant mass distribution
and forward-backward asymmetry for $\ell=e, \mu, \tau$, and the tau
polarization asymmetry for \bstt.  A three
dimensional $\chi^2$ fit to the coefficients $C_{7,9,10}(\mu)$ is 
performed for three values of the integrated luminosity, 
$3\times 10^7$, $10^8$, and $5\times 10^8$ $B\bar B$ pairs,
corresponding to the expected $e^+e^-$ $B$-factory luminosities of one
year at design, one year at an upgraded accelerator, and the total
accumulated luminosity at the end of the programs.  
The $95\%$ C.L. allowed regions as projected
onto the $C_9(\mu)-C_{10}(\mu)$ and $C^{eff}_7(\mu)-C_{10}(\mu)$ planes are
depicted in Figs.~\ref{fig3}(a-b), where the diamond
represents the central value for the expectations in the SM.  
We see that the determinations are
relatively poor for $3\times 10^{7}$ $B\bar B$ pairs and that
higher statistics are
required in order to focus on regions centered around the SM.  

We analyze the supersymmetric contributions
to the Wilson coefficients in terms of the quantities
\beq
R_i\equiv \frac{C^{susy}_i(M_W)}{C^{SM}_i(M_W)}-1\equiv 
{C_i^{new}(M_W)\over C_i^{SM}(M_W)}\,,
\eeq
where $C^{susy}_i(M_W)$ includes the full SM plus superpartner
contributions.  $R_i$ is meant to indicate the relative fraction difference
from the SM value.  

Supersymmetry has many potential sources for flavor violation.  The
flavor mixing angles among the squarks are {\it a priori} 
separate from the CKM angles of the SM quarks.  
We adopt the viewpoint here that 
flavor-blind (diagonal) soft terms~\cite{dimo2} 
at the high scale are the phenomenological
source for the soft scalar masses, and that the CKM
angles are the only relevant flavor violating sources.  The spectroscopy of 
the supersymmetric states is quite model dependent and here we 
analyze two possibilities.  The first is the familiar
minimal supergravity model.  In this case all the
supersymmetric states follow from a common scalar mass 
and a common gaugino mass at the high scale.  The second possibility is to 
relax the condition of common scalar masses and allow
them to take on uncorrelated values at the low scale
while still preserving gauge invariance.

\begin{figure}[t]
\centerline{
\psfig{figure=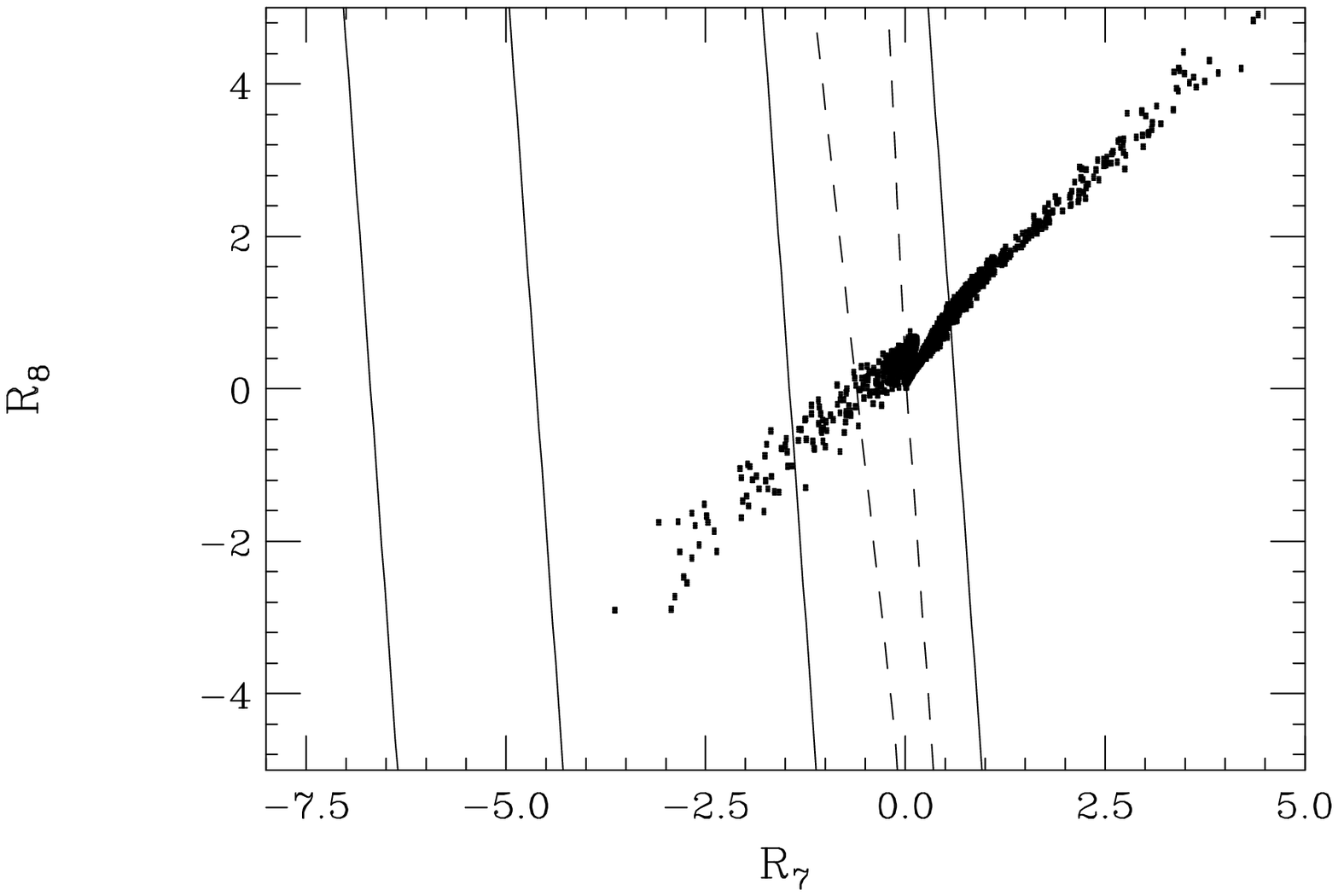,height=1.75in,width=2.5in}}
\centerline{
\psfig{figure=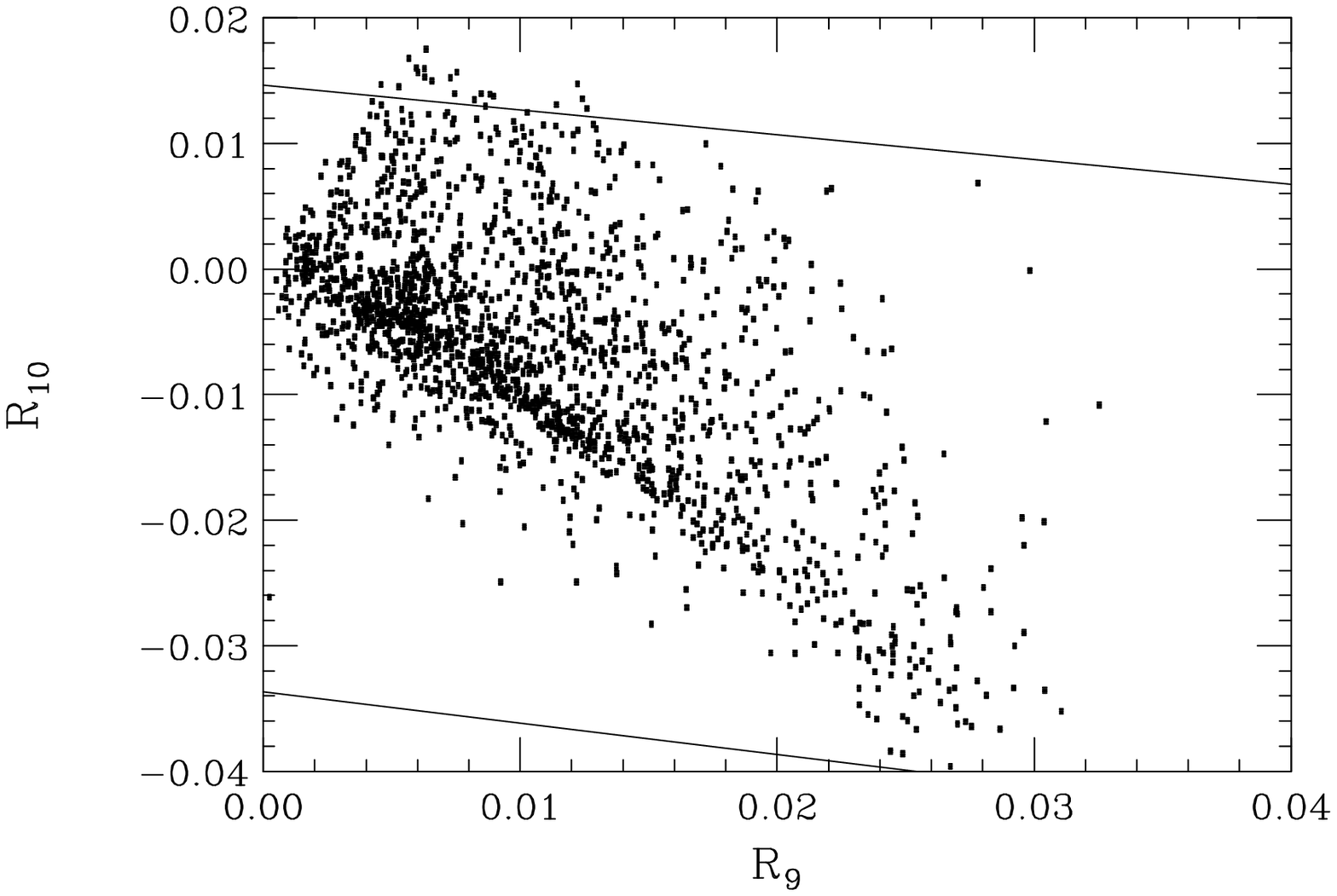,height=1.75in,width=2.5in}}
\vspace*{-0.25cm}
\caption{(a) Parameter 
space scatter plot of $R_7$ vs. $R_8$ in the minimal supergravity model.
The allowed region from CLEO data lies inside the
solid diagonal bands.  The dashed band represents the potential
$10\%$ measurement from the previously described global fit to the
coefficients. (b) Parameter 
space scatter plot of $R_9$ 
vs. $R_{10}$. The global fit to the coefficients
obtained with $5\times 10^8\, B\bar B$ pairs
corresponds to the region inside the diagonal bands.}
\label{fig4}
\end{figure}

We begin by searching over the full parameter space of the minimal supergravity
model.  We generate \cite{kkrw} these models by applying common soft scalar 
and common gaugino masses at the boundary scale.
The tri-scalar $A$ terms
are also input at the high scale and are universal.
The radiative electroweak symmetry 
breaking conditions yield the $B$
and $\mu^2$ terms as output, with a $\mbox{sign}(\mu )$ ambiguity left over.
(Here $\mu$ refers to the Higgsino mixing
parameter.) We also choose $\tan\beta$ and restrict it
to a range which will yield perturbative Yukawa couplings up to the GUT scale.
We have generated thousands of solutions according to the above procedure.  The
ranges of our input parameters are $0< m_0 < 500\gev$, $50< m_{1/2} < 250\gev$,
$-3 < A_0/m_0 < 3$, $2 < \tan\beta < 50$,
and we have taken $m_t=175\gev$.  Each supersymmetric solution is
kept only if it is not in violation with present constraints from
SLC/LEP and Tevatron direct sparticle production limits, and it is out of reach
of LEP II.  For each of these remaining solutions 
we now calculate $R_{7-10}$.
Our results are shown in the scatter plots of Fig.~\ref{fig4}
in the (a) $R_7-R_8$  and (b) $R_9-R_{10}$ planes.  The diagonal
bands represent the bounds on the Wilson coefficients as previously
determined from our global fit.  We see that the current CLEO data on \bsg\
already places signigicant restrictions on the supersymmetric parameter space.

The first thing to note from the figure
is that large values of $R_7$ and $R_8$ are generated, and that they
are very strongly correlated.   
These large effects arise from models with $|\mu |\lsim 400\gev$. 
This is because light charged Higgsinos (or rather light charginos with a
large Higgsino fraction) are necessary in order to obtain a large effect
on the Wilson coefficients.  
We see that the values of $R_9$ and $R_{10}$ are bounded by
about $0.04$, which is small compared to the range for $R_7$.  
The main reason for
these smaller values is the dependence on the bottom Yukawa 
$\lambda_b\propto 1/\cos\beta$.  $R_7$ also has a contribution directly
dependent on this $1/\cos\beta$ Yukawa factor, however
the other multiplicative
terms associated with $\lambda_b$ are the large top Yukawa and a
large kinematic loop factor.  $R_{9}$ and $R_{10}$ do not have such 
additional factors due the chirality structure of these operators and the
requirement that leptons and sleptons only couple via $SU(2)$ and $U(1)$
gauge couplings.  These conditions, 
along with the correlations between the mass spectra dictated by
minimal supergravity relations, render the minimal supergravity 
contributions to
$R_{9,10}$ essentially unobservable.  

We now adopt our second, more phenomenological, approach.
The maximal effects for the parameters $R_i$ can be estimated
for a superparticle spectrum independent of the high scale assumptions.
However, we still maintain the assumption that CKM angles alone constitute
the sole source of flavor violations in the full supersymmetric lagrangian.
We will focus on the region 
$\tan\beta \lsim 30$.
The most important features which result in large contributions are
a light $\tilde t_1$ state present in the SUSY spectrum and at least one
light chargino state.  For the dipole moment operators a light Higgsino
is most important.  A pure higgsino and/or pure
gaugino state have less of an effect than two mixed states when searching
for maximal effects in $C_9$ and $C_{10}$ and 
we have found that $M_2 \simeq 2 \mu$ is optimal.

\begin{figure}[t]
\centerline{
\psfig{figure=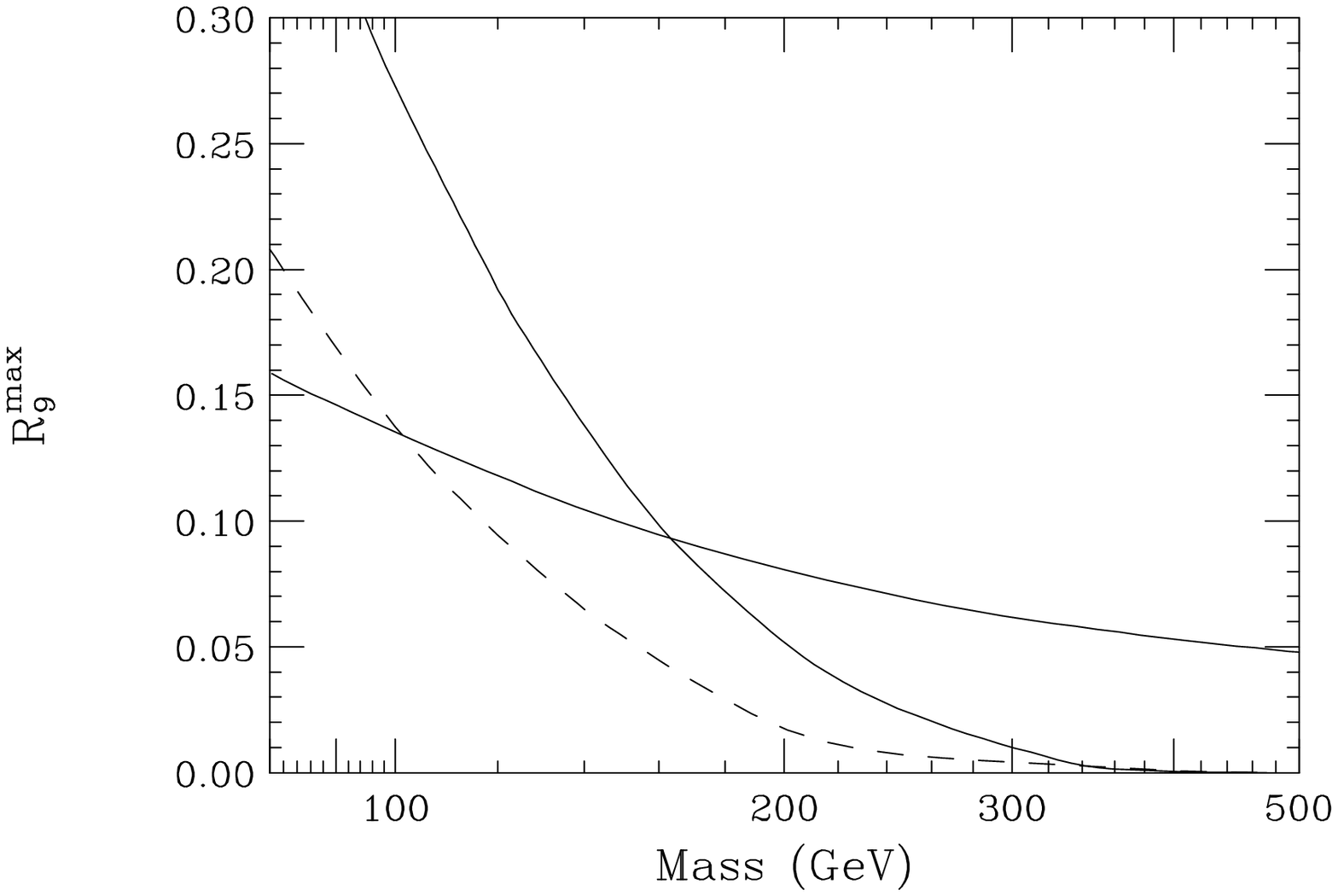,height=1.75in,width=2.5in}}
\centerline{
\psfig{figure=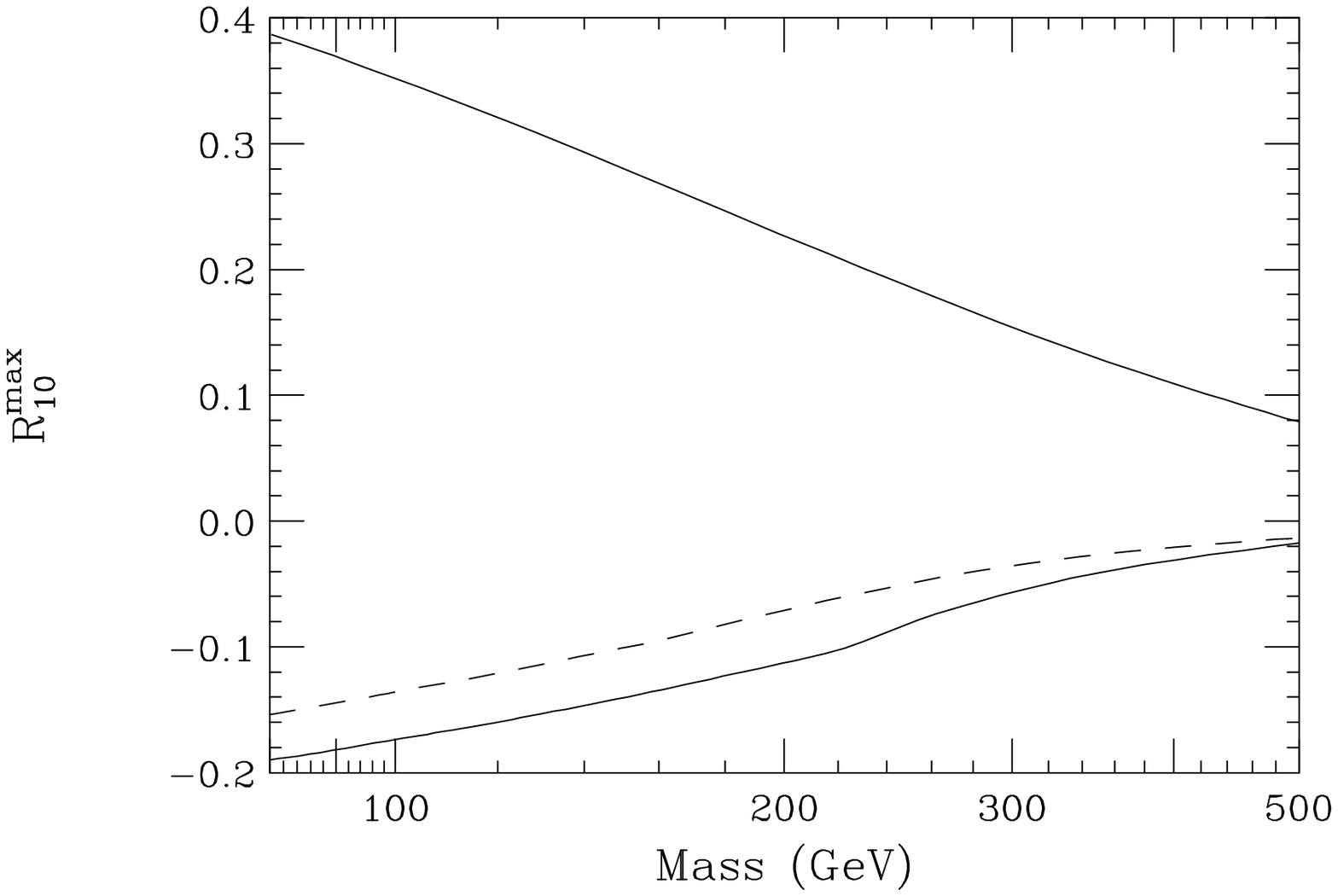,height=1.75in,width=2.5in}}
\vspace*{-0.25cm}
\caption{The maximum value of (a) $R_9$ and
(b) $R_{10}$
achievable for general 
supersymmetric models. The top solid line comes from $t-H^\pm$
contribution and is displayed versus the $H^\pm$ mass.  The bottom solid line
is from $\tilde t_i -\chi^\pm_j$ contribution with $\tan\beta =1$
and is shown versus the $\chi^\pm_i$ mass.  The dashed line is
the $\tilde t_i -\chi^\pm_j$ contribution with $\tan\beta =2$.
The other mass parameters which are not plotted are chosen to be just above
LEPII and Tevatron's reach.}
\label{fig8}
\end{figure}

Fig.~\ref{fig8} displays the
maximum contribution to $R_{9,10}$
versus an applicable SUSY mass scale.  
The other masses which are not plotted ($\tilde t_i$,
$\tilde l_L$, etc.) are chosen to be just above the reach of LEPII or
the Tevatron, whichever gives better bounds.  We see that the maximum 
size of $R_{9,10}$ is much larger than what was allowed in 
the minimal supergravity model. This is due to the lifted restriction
on mass correlations.  Light
sleptons, sneutrinos, charginos, and stops are allowed simultaneously
with mixing angles giving 
the maximal contribution to the $R_i$'s.  However, we find that the
maximum allowed values for $R_{9,10}$ are still much less than unity.  
Earlier we determined that $B$ factory data would be sensitive to
$\Delta R_9\gsim 0.3$ and $\Delta R_{10}\gsim 0.08$ at the highest 
luminosities, 
and so the largest SUSY effect would give a
$~1-2\sigma$ signal in $R_{9,10}$, hardly enough to be a compelling indication
of physics beyond the standard model.  

Given the sensitivity of all the 
observables it is instructive to narrow the focus to $C_7(M_W)$.  
There exists the possibility that one
eigenvalue of the stop squark mass matrix might be much
lighter than the other squarks due to the large top Yukawa and the
mixing term $m_t(A_t-\mu\cot\beta)$ in the stop mass matrix.  
We then present results for $C_7(M_W)$ in the limit of
one light squark, namely the $\tilde t_1$, and light charginos.  
We allow the $\tilde t_1$ to have
arbitrary components of $\tilde t_L$ and $\tilde t_R$ since cross terms
can become very important.  This is especially noteworthy 
in the high $\tan\beta$ limit.  
We note that the total supersymmetric contribution to $C_7(\mw)$ 
will depend on several combinations of
mixing angles in both the stop and chargino mixing matrices and
cancellations can occur for different signs of $\mu$~\cite{garisto93:372}.  

\begin{figure}
\centerline{
\psfig{figure=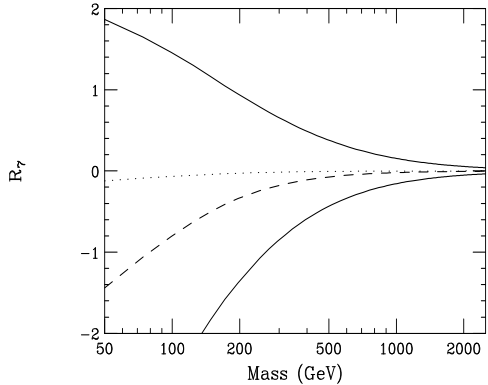,height=2.0in,width=2.5in}}
\caption{Contributions 
to $R_7$ in the different limits described in the text.  The top solid line
is the charged $H^\pm/t$ contribution versus $m_{H^\pm}$.  
The bottom solid line is the
$\tilde \chi^\pm_1/\tilde t_1$ contribution versus $m_{\tilde \chi^\pm}$
where both 
the chargino and stop are maximally
mixed states with $\mu <0$.   
The dashed line is the $\tilde H^\pm /\tilde t_R$ contribution,
and the dotted line represents the $\tilde W^\pm /\tilde t_1$ contribution.
These two lines are both shown as a function of 
$\tilde \chi^\pm_1$ mass.
All lines are for $\tan\beta =2$ and $m_t=175\gev$.
We have set all other masses to be just above the reach of LEPII.}
\label{fig11}
\end{figure}

The first case we examine is that where the lightest chargino is
a pure Higgsino and the lightest stop is purely right-handed:  
$\ca \sim \tilde H^\pm$, $\tilde t_1 \sim \tilde t_R$.  
The resulting contribution to $R_7$ is shown
as a function of the $\tilde t_R$ mass in Fig.~\ref{fig11} (dashed line) for
the case of chargino masses out of reach of LEP II ($m_{\ca} \gsim M_W$).
Note that the SUSY contribution to $C_7(M_W)$ in this limit always adds
constructively to that of the SM.
Next we examine the limit where the only light chargino is a pure Wino.
The effects of a light pure Wino are generally
small since (i) it couples with gauge strength rather than the top 
Yukawa, and (ii) 
generally supersymmetric models do not yield a light $\tilde t_L$ necessary
to couple with the Wino.  
This contribution 
to $R_7$ is shown in Fig.~\ref{fig11} (dotted line).  As expected, we
see that this contribution is indeed small. 
Now we discuss our third limiting case.
As mentioned above, what we mostly expect in minimal supergravity models
is a highly mixed $\tilde t_1$ state; here, for large effects, 
it is crucial that there be substantial $\tilde t_R$ and $\tilde t_L$ 
contributions to $\tilde t_1$.  We find that in this case
large $\tan\beta$ solutions ($\tan\beta \gsim 40$)
can yield greater than ${\cal O}(1)$ contributions to
$R_7$ even for SUSY scales of $1\tev$! Low values of $\tan\beta$ can also
exhibit significant enhancements.  This
is demonstrated for $\tan\beta =2$ in Fig.~\ref{fig11} (solid line).
Note that in this case large contributions are
possible in both the negative and positive directions of $R_7$ depending on the
sign of $\mu$.  We note that this is a region of SUSY parameter space
which is highly motivated by $SO(10)$ grand unified theories.

In this talk we have studied the effects of supersymmetry
to the FCNC observables concerning $b\to s$ transitions, 
and we have seen that deviations from
the standard model could be detected with supersymmetric masses 
even at the TeV scale.  This is especially true if $\tan\beta$ is very high.
The large $\tan\beta$ enhancements in the $b\to s$ processes are unique;
it is thus possible that the first distinct signs
of supersymmetry could come from deviations in rare $B$ decays. 
One, of course, would like direct confirmation of such a deviation, if
observed at $B$-Factories, and collider programs could provide it.  


\begin{thebibliography}{20}

\bibitem{cleo:94} CLEO Collaboration, M.S. Alam \etal, \PRL 74 2885 1995 ;
CLEO Collaboration, R. Ammar \etal, \PRL 71 674 1993 .

\bibitem{bsll} CLEO Collaboration, R.~Balest \etal, in {\it Proceedings
of the 27th Int. Conf. on HEP}, Glasgow,
Scotland, 1994, edited by P.J.~Bussey and I.G.~Knowles (IOP, London,
1995); CDF Collaboration, C.~Anway-Wiese, in {\it The Albuquerque Meeting},
Proceedings of the 8th Meeting of the DPF of the APS,
Albuquerque, New Mexico, 1994, edited by S.~Seidel
(World Scientific, Singapore, 1995).

\bibitem{ali} A. Ali, T. Mannel, and T. Morozumi, \PLB B273 505 1991 ;
A. Ali, G.F. Giudice, and T. Mannel, \ZPC C67 417 1995 .

\bibitem{hewett} J.L.~Hewett, \PRD D53 4964 1996 .

\bibitem{jimjo} J.L. Hewett and J.D. Wells, SLAC-PUB-7290, hep-ph/9610323.

\bibitem{rareb}
S. Bertolini, F.~Borzumati, A.~Masiero, G.~Ridolfi, \NPB B353 591 1991 .
F. Borzumati, \ZPC C63 291 1994 ;
F. Gabbiani, E. Gabrielli, A. Masiero, L. Silvestrini, hep-ph/9604387;
V. Barger, M.S. Berger, P. Ohmann, R.J.N. Phillips, \PRD D51 2438 1995 ;
D. Choudhury, F. Eberlein, A. Konig, J. Louis, S. Pokorski, \PLB B342 180 1995 ;
J. Lopez, D. Nanopoulos, X. Wang, A.~Zichichi, \PRD D51 147 1995 .
R. Barbieri, G. Giudice, \PLB B309 86 1993 ;
P. Cho, M. Misiak, D. Wyler, hep-ph/9601360.

\bibitem{buras:95} A.J. Buras and M. M\" unz, \PRD D52 186 1995 .

\bibitem{inami} T. Inami and C.S. Lim, Prog. Theor. Phys. {\bf 65}, 297
(1981).

\bibitem{pdg} R.M. Barnett, \etal, (Particle Data Group), \PRD D54 1 1996 .

\bibitem{schmelling} M.~Schmelling, these proceedings.

\bibitem{cdfd0} P. Tipton, these proceedings.

\bibitem{richman} J. Richman, these proceedings.

\bibitem{greub:91} A. Ali and C. Greub, \ZPC C49 431 1991 ; 
\PLB B259 182 1991 ; \PLB B361 146 1995 ; N. Pott, \PRD D54 938 1996 .

\bibitem{greub:96} C. Greub, T. Hurth, and D. Wyler, \PLB B380 385 1996 ;
\PRD D54 3350 1996 .

\bibitem{yao} K. Adel and Y.-P. Yao, \PRD D49 4945 1994 .

\bibitem{misiak:96} K.G. Chetyrkin, M. Misiak, and M. M\" unz, in preparation;
M. Misiak, these proceedings. 

\bibitem{dimo2} S.~Dimopoulos, H.~Georgi, \NPB B193 150 1981 .

\bibitem{kkrw}For description of the procedure we follow, see 
G.L.~Kane, C.~Kolda, L.~Roszkowski, J.~Wells, \PRD D49 6173 1994 .

\bibitem{garisto93:372}R. Garisto, J. Ng, \PLB B315 372 1993 .

\end{thebibliography}
\end{document}